 \documentclass[pdflatex,sn-nature,iicol]{sn-jnl}

\usepackage{array}
\usepackage{graphicx}%
\usepackage{multirow}%
\usepackage{amsmath,amssymb,amsfonts}%
\usepackage{amsthm}%
\usepackage{mathrsfs}%
\usepackage[title]{appendix}%
\usepackage{xcolor}%
\usepackage{textcomp}%
\usepackage{manyfoot}%
\usepackage{booktabs}%
\usepackage{algorithm}%
\usepackage{algorithmicx}%
\usepackage{algpseudocode}%
\usepackage{listings}%
\usepackage{setspace}
\usepackage[switch]{lineno}
\usepackage{geometry}
\usepackage[T1]{fontenc}

\theoremstyle{thmstyleone}%
\theoremstyle{thmstyletwo}%

\theoremstyle{thmstylethree}%

\begin{document}

\setstretch{1.0}

\title[Article Title]{Community and hyperedge inference in multiple hypergraphs}



\author[1,2]{ \fnm{Li} \sur{Ni}}\email{nili@ahu.edu.cn}

\author[1]{\fnm{Ziqi}  \sur{Deng}}\email{e23201029@stu.edu.ahu.cn}

\author[1]{\fnm{Lin}  \sur{Mu}}\email{mulin@ahu.edu.cn}

\author[1]{\fnm{Lei} \sur{Zhang}}\email{zl@ahu.edu.cn}

\author*[3]{\fnm{Wenjian}  \sur{Luo}}\email{luowenjian@hit.edu.cn}

\author*[1]{\fnm{Yiwen} \sur{Zhang}}\email{zhangyiwen@ahu.edu.cn}

\affil*[1]{\orgdiv{School of Computer Science and Technology}, \orgname{Anhui University}, \orgaddress{ \city{Hefei}, \postcode{230601}, \state{Anhui}, \country{China}}}

\affil[2]{\orgdiv{Key Laboratory of Intelligent Computing \& Signal Processing}, \orgname{Ministry of Education}, \orgaddress{\city{Hefei}, \postcode{230601}, \state{Anhui}, \country{China}}}

\affil*[3]{\orgdiv{School of Computer Science and Technology}, \orgname{Harbin Institute of Technology}, \orgaddress{ \city{Shenzhen}, \postcode{518055}, \country{China}}}


\abstract{
Hypergraphs, capable of representing  high-order interactions via hyperedges, have become a powerful tool for modeling real-world biological and social systems.
Inherent relationships within these real-world systems, such as  the encoding relationship between genes and their protein products, drive the establishment of interconnections between multiple hypergraphs. 
Here, we demonstrate how to utilize those interconnections between multiple hypergraphs to synthesize integrated information from multiple higher-order systems, thereby enhancing  understanding of underlying structures.
We propose a model based on the stochastic block model, which integrates information from multiple hypergraphs  to reveal latent high-order structures.
Real-world hyperedges exhibit preferential attachment, where certain nodes dominate hyperedge formation. To characterize this phenomenon,  our model introduces hyperedge internal degree to quantify nodes' contributions to hyperedge formation. 
This model is capable of  mining communities, predicting missing hyperedges of arbitrary sizes within hypergraphs, and inferring inter-hypergraph edges between hypergraphs.
We apply our model to  high-order datasets to evaluate its performance.
Experimental results demonstrate strong performance of our model in community detection, hyperedge prediction, and inter-hypergraph edge prediction tasks. 
Moreover, we show that our model enables analysis of  multiple hypergraphs of different types and supports the analysis of a single hypergraph in the absence of inter-hypergraph edges. 
Our work provides a practical and flexible tool for analyzing multiple hypergraphs, greatly advancing the understanding of the organization in real-world high-order systems.
} 




\maketitle

\section{Introduction}\label{sec1}
High-order interactions among three or more system units have been observed in various systems, such as collaboration networks \cite{bib2}, cellular networks \cite{bib1}, and drug recombination \cite{bib3}. 
Modeling networks with such high-order interactions requires advanced mathematical frameworks.
Among these frameworks, hypergraphs are an emerging paradigm for modeling these high-order networks \cite{para1,para1-1,para1-2}, whose hyperedges of arbitrary sizes can effectively capture the structured relations among multiple system units \cite{bib5,bib6,bib7}.  
So far, several techniques for analyzing hypergraph emerges, such as high-order centrality analysis \cite{bib8,bib9}, link prediction \cite{bib10}, and community detection \cite{bib11}. 
Among these, community detection is widely applied in various fields, such as parallel computing \cite{bib15,bib16}, circuit design \cite{bib17}, and image segmentation \cite{bib18}.
To detect communities in hypergraphs, researchers propose numerous methods, including tensor decompositions \cite{bib13}, flow-based algorithms \cite{bib22,bib23}, spectral clustering-based approaches \cite{bib26,Spectral_Method,Spectral_Method2}, and methods based on statistical inference \cite{bib10,bib20}.  
Among these, methods based on statistical inference possess a solid theoretical foundation in mathematics and exhibit high computational efficiency \cite{bib19,bib21,bib24,bib34-3x}. 

The above  studies  \cite{bib34-3x,bib35-3x,bib36-3x} focus on the exploration of the single high-order system, and their performance is limited by the information of a single hypergraph.
 %
However, multiple high-order real world systems  are interconnected \cite{bib30-3x,bib31-3x}, such as the coding relationship between genes in  the gene co-expression hypergraph and proteins in the protein interaction  hypergraph \cite{bib27-3x,bib28-3x,bib29-3x}.
This situation aligns with findings from  work \cite{para2biological,para2biological2}, which emphasized the necessity of multi-network frameworks by integrating gene co-expression networks, protein interaction networks, and metabolic pathways to reconstruct comprehensive biomolecular systems \cite{para2biological,para2biological2}. 
Therefore, we explore the incorporation of information from multiple hypergraphs for detecting community structures, thereby capturing complex multi-level interactions and high-order structural features within the multiple systems. 


A key aspect of exploring multiple hypergraphs is the modeling of hyperedges. 
Existing approaches model hyperedges by  treating all nodes within a hyperedge as equally important.
However, our observation of preferential attachment \cite{preferential-attachment-mechanism} in hyperedges, where certain nodes consistently attract more connections, directly implies the  intrinsic difference in node importance.
This differential importance is clearly manifested in cases like meeting hyperedges, where the host’s role is pivotal for steering discussions, while the audience primarily contributes through passive engagement.
Therefore,  we  consider the differential importance of nodes in modeling hyperedges. Specifically, we introducing the internal degree of hyperedges in hyperedge modeling to quantify  the contributions of nodes to the generation of hyperedges.


To this end, to model multiple related high-order systems, we propose a method following the generative modeling framework \cite{bib36-3x, generative_model}.
Based on the principles of generative models, we infer latent variables from the observed interactions within hypergraphs and between hypergraphs to capture the community structure.
The proposed model exhibits two desirable features. 
On the one hand,  it supports high-order systems with identical or different node types, making it applicable to multi-layer and multi-domain hypergraphs.
On the other  hand, the model allows flexible adjustment of node contributions to hyperedge generation  according to actual circumstances, thereby enhancing the model's adaptability to complex network structures. 

By fully considering interactions among hypergraphs, the model integrates information from real-world high-order systems to provide comprehensive insights into complex multiple systems.
Compared with single-hypergraph structures, multi-hypergraph structures more accurately characterize the complex interdependencies among higher-order interactive systems in real-world networks \cite{para-last-multilayer, para-last-multilayer2}. Our model enables systematic analysis of multi-hypergraphs, thereby providing novel theoretical frameworks and methodological tools for understanding and modeling behavioral patterns in multiple higher-order systems.


\section{Results}\label{sec2}
\subsection{The model}\label{subsec1}
We denote a multi-hypergraph as $ MH = \{ H^1, \dots,$ $ H^l,  \dots, H^L, S^{l l^{\prime}} \} $, where  $H^l = ( V^l, E^l)$ represents the $l$-th hypergraph, \( S^{l l^{\prime}} = \{ S_{ij}^{l l^{\prime}}|i\in V^l,j\in V^{l^\prime} , l,l^{\prime} \in [0,L] \} \) is the set of inter-hypergraph edges between hypergraphs, and  $L$ means the number of hypergraphs.
For \( H^l = ( V^l, E^l) \), \(  V^l = \{ v_1^l, \dots, v_k^l \} \) represents  a set of nodes  and  \( E^l = \{ e_1^l, \dots, e_n^l \} \)  represents  a hyperedge set.
Each hyperedge \( e_i^l \) is a subset of the node set \(  V^l \), denoting a high-order interaction among the \( |e_i^l| \) nodes. 
In this paper, the structure of hypergraph \( H^l \) is represented by the adjacency vector \( A^l \in \mathbb{N}^{\Omega^l} \), where \( A_e^l \in A^l \) denotes the weight of the hyperedge \( e^l\in \Omega^l \) and
\( \Omega^l \) is the set of all possible hyperedges among the nodes in \(  V^l \).
A non-zero value of \( A_e^l \) indicates the presence of the corresponding hyperedge \( e \), while a zero value indicates that the hyperedge does not exist.


Here, we propose a Multi-Hypergraph Stochastic Block Model (MHSBM). Grounded in statistical inference principles, it simultaneously models both hyperedges within hypergraphs and inter-hypergraph edges between hypergraphs.
MHSBM leverages complementary information from different hypergraphs to improve the completeness of information.
Assuming  the hyperedge weights  $A = \{A^1,A^2,\dots, A^L  \}$ of $L$ hypergraphs and inter-hypergraph edges $S=\{S^{12},\cdots,S^{l l^{\prime}}\}$ between hypergraphs are conditionally independent,  we model  $A$ and $S$ by a set of latent variables 
$\Phi = (\{u^1,\dots, u^l,\dots,u^L\},\{w^1,\dots,w^l,\dots ,w^L\},\{w^{ll^{\prime}}| $ $ S^{ll^{\prime}}\neq    \emptyset , l,l^{\prime} \in [0,L] \})$, as follows:
\begin{equation}
\begin{aligned}
\small
   & P(A,S|\Phi)= \\
    &\prod_{l \in [0,L]}P_A(A^l|u^l, w^l) 
    \prod_{l^{\prime} \ne l;l,l^{\prime} \in [0,L]}P_S(S^{l l^{\prime}}|u^l,u^{l^{\prime}},w^{l l^{\prime}}),  \label{eq1}
\end{aligned}
\end{equation}
where $u^l$ represents the node-community membership matrix of the $l$-th hypergraph, $w^l $ denotes the community affinity matrix of the $l$-th hypergraph, and $w^{ll^{\prime}} $ indicates the inter-hypergraph community affinity matrix between the $l$-th and $l^{\prime}$-th hypergraphs.
The term $P_A(A^l|u^l, w^l)$  models the  $l$-th hypergraph, while the term $P_S(S^{l l^{\prime}}|u^l,u^{l^{\prime}},w^{l l^{\prime}})$ represents the modeling of inter-hypergraph edges between $l$-th hypergraph and $l^{\prime}$-th hypergraph.

Having introducted the model structure, we present a detailed introduction of the $P_A(A^l|u^l, w^l)$  and $P_S(S^{l l^{\prime}}|u^l,u^{l^{\prime}},w^{l l^{\prime}})$ in Section \ref{subsubsec1} and Section \ref{subsubsec2}, respectively.

\subsubsection{Modeling hyperedges in hypergraphs } \label{subsubsec1}
We propose an improved version of the hypergraph stochastic block model (Hy-MMSBM) \cite{bib19}, thereby enhancing its adaptability to complex real-world system.
The Hy-MMSBM model treats all nodes within a hyperedge as equally important to the generation of a hyperedge. 
However, in practical scenarios, nodes often have different strength of influence on the generation of hyperedges.
Taking workplace scenarios as an example, the hyperedge represents meeting participation sets, where managers usually make greater contributions than employees in facilitating the formation of hyperedges for meetings.
To modulate node contributions to formation of hyperedges, we design the hyperedge internal degree  to control the importantance of nodes in hyperedges. 



We assume that the \(l\)-th hypergraph (\(l \in [0,L]\)) contains \(K^l\) latent communities.
The $K^l \times K^l$ community affinity matrix $w^l$ regulates the interaction strength between communities within the hypergraph.
$N^l \times K^l$ membership matrix $u^l$ describes the membership of nodes belongs to multiple communities.
$N^l \times E^l$ hyperedge internal degree matrix $\theta^l$  controls the contribution of each node to the hyperedges where  $\theta_{ie}^{l}$ represents the contribution of node $v_i$ to hyperedge $e$.
Assuming hyperedges follow a Poisson distribution, parameterized by the given $u^l$, $w^l$, and $\theta^l$, the probability of hyperedges in the $l$-th hypergraph is modeled as follows:
    \begin{equation}
    \begin{split}
        P_A(A^l|u^l, w^l, \theta^l)  =\prod_{e \in \Omega^l}Pois(A_{e}^l;\frac{\lambda^{l}_{e}}{\mu_e^l}) ,\label{eq2} 
          \end{split}
    \end{equation}
where
    \begin{equation}
        \begin{gathered}
        \lambda_{e}^{l}=\sum_{i<j:i,j\in e} \theta_{ie}^l  u_{i}^{l} w^{l} ~\theta_{je}^l ({u_{j}^{l}})^T\\
        =\sum_{i < j \in e} \sum_{k,q=1}^K \theta_{ie}^l \theta_{je}^l u_{ik}^l u_{jq}^l w_{kq}^l \label{eq3},
        \end{gathered}
    \end{equation} \( \mu_e^l \) is a normalization constant. 
For hyperedge $e$ of size \( n \), we calculate the normalization constant using the form \( \mu_e = \frac{n(n-1)}{2} \), where \( \frac{n(n-1)}{2} \) represents the number of possible pairwise interactions among the \( n \) nodes in the hyperedge.

Hyperedge internal degree $\theta_{ie}$ denotes the degree of node $i$ within hyperedge $e$, which describes the contribution of node $i$ to the hyperedge e. It is formulated as:
\begin{equation}
    \theta_{ie} = |e| \cdot \frac{\varepsilon _{ie}}{\Sigma_{j\in e} \varepsilon _{je} },
\end{equation}
where $\varepsilon_{ie} = |\{e^{\prime} \mid i\in e^{\prime}, e^{\prime} \subseteq e:e^{\prime} \in E \}|$ denotes the number of sub-hyperedges within hyperedge $e$ that contain node $i$, and $\sum_{j\in e}\varepsilon _{je}$ normalizes the numerator. 
$|e|$ represents the size of hyperedge $e$, and ensure that the sum of the hyperedge internal degree of all nodes within the hyperedge equals its size.
The hyperedge internal degree $\theta_{ie}$ adjusts node $v_i$'s importance in hyperedge $e$ by quantifying its participation through $\varepsilon_{ie}$. 

\subsubsection{Modeling inter-hypergraph edges}\label{subsubsec2}
Similar to the hyperedge modeling way in Section \ref{subsubsec1}, we employ a probabilistic approach for modeling the interaction edges between hypergraphs. 
The emergence of interactions arises from the association between community structures across multiple hypergraphs, i.e., topic-based paper communities in citation networks correspond to domain-specific scholar communities in collaboration networks. 
Such across hypergraphs association is reflected in the proximity of membership distributions.
Therefore, the inter-hypergraph edges are modeled on the basis of membership vector \( u^l \) of the $l$-th hypergraph and \( u^{l'} \) of the $l^{\prime}$-th hypergraph.
Since the association strength between between communities in the  $l$-th hypergraph and those in the  $l^{\prime}$-th hypergraph may be strong or loose,  it is  controlled by a non-negative affinity matrix \( w^{ll'} \) of dimensions \( K^l \times K^{l'} \) (\( l, l^{\prime} \in [0,L] \)). Here,  \( K^l \) and \( K^{l'} \) represent the number of communities in $l$-th hypergraph and  the  $l^{\prime}$-th hypergraph, respectively.
We assume that the edges between hypergraphs are conditionally independent and follow Poisson distributions.
The Poisson distribution for inter-hypergraph edges are then determined jointly by parameters \( u^l \), \( u^{l'} \), and \( w^{ll'} \): 
\begin{equation}
\begin{split}
   P_S(S^l|u^l, u^{l^{\prime}},w^{l l^{\prime}})
   &=  \prod_{\substack{i \in V^l, j\in V^{l^{\prime}}}} \text{Pois}(S_{ij}^{ll^{\prime}}; \lambda_{ij}^{ll^{\prime}}), \label{eq4}
\end{split}
\end{equation}
where $\lambda_{ij}^{ll^{\prime}}$ is given by

\begin{equation}
    \begin{gathered}
\lambda_{ij}^{ll^{\prime}} = u_{i}^{l} w^{ll^{\prime}} {u_{j}^{l^{\prime}}}^T = \sum_{k,c} u_{ik}^{l} w_{kc}^{ll^{\prime}} u_{jc}^{l^{\prime}}.  \label{eq5}
  \end{gathered}
\end{equation}


Note that existing community detection models   \cite{bib10,bib19} utilize only information from a single hypergraph  to detect communities.
In contrast, our model combines information from multiple hypergraphs to detect communities. In our model, inter-hypergraph affinity matrix \(w^{ll^{\prime}}\) quantifies the strength of interactions between the communities in the \(l\)-th and \(l^{\prime}\)-th hypergraphs. 
This allows  our model to learn the interactions between communities across hypergraphs, thereby providing complementary information to capture high-order structural information. 
Since no prior constraints are imposed on $w^{ll^{\prime}}$, 
it makes our model adaptable to complex high-order multiple hypergraphs with distinct or identical node types. 
\subsubsection{Variable inference }\label{subsubsec3}

Based on defined Eq. (\ref{eq1}) and the observed  multi hypergraphs, we consider maximum likelihood estimation to infer $\Phi = (\{u^1,\dots, u^l,\dots,u^L\},\{w^1,\dots,w^l,\dots ,w^L\},\{w^{ll^{\prime}}| $ $ S^{ll^{\prime}}\neq    \emptyset , l,l^{\prime} \in [0,L] \})$. 
The latent variables  are iteratively optimized through the Expectation-Maximization (EM) algorithm \cite{EM}, as detailed in the Methods section.

\subsection{Results on real datasets}\label{subsec3}

We adopted real-world hypergraphs \cite{bib37-highschooldatabase,bib38-primaryschool,workplace,hospital,senatebill} from diverse domains to evaluate the performance of models.  
To simulate multiple hypergraph scenarios, we constructed two hypergraphs by independently sampling either 70\% or 80\% of hyperedges  twice, labeled H0 and H1, repectively. 
Subsequently, inter-hypergraph edges were generated between the hypergraphs based on their community structures. Noise edges were also added.
Furthermore, to validate the model's performance on multi-domain hypergraph, 
we constructed two multi-domain hypergraphs using biological data and academic network information respectively.
Table \ref{tab:datasets} reports information of multi-hypergraphs.
.


\begin{table}[ht]
\setlength{\tabcolsep}{7pt}
\centering
\caption{Statistics on higher-order datasets}
\label{tab:datasets}
\begin{tabular}{p{0.28\columnwidth}p{0.09\columnwidth}p{0.07\columnwidth}p{0.1\columnwidth}p{0.04\columnwidth}p{0.03\columnwidth}}
\toprule
Dataset Name & Layer & N & |E| & D & K \\
\midrule
\multirow{3}{*}{Highschool} & H   & 327 & 7818 & 5 & 9 \\
                            & H0  & 327 & 6254 & 5 & 9 \\
                            & H1  & 327& 6255 & 5 & 9 \\
\cmidrule{1-6}
\multirow{3}{*}{Workplace}  & H   & 92  & 788  & 4 & 5 \\
                            & H0  & 92  & 551  & 4 & 5 \\
                            & H1  & 92  & 552  & 4 & 5 \\
\cmidrule{1-6}
\multirow{3}{*}{Hospital}   & H   & 75  & 1825 & 5 & 4 \\
                            & H0  & 75  & 1277 & 5 & 4 \\
                            & H1  & 75  & 1278 & 5 & 4 \\
\cmidrule{1-6}
\multirow{3}{*}{Primaryschool}& H   & 242 & 12704& 5& 11\\
                            & H0  & 242  & 8892 & 5 & 11 \\
                            & H1  & 242  & 8894 & 5 & 11 \\
\cmidrule{1-6}
\multirow{3}{*}{House\_bills}& H   & 1494 & 54933& 399& 2\\
                            & H0  & 1494  & 43946 & 399 & 2 \\
                            & H1  & 1494  & 43947 & 399 & 2 \\
\cmidrule{1-6}
\multirow{3}{*}{House\_committees}& H   & 1245 & 335& 81& 2\\
                                    & H0  & 1245  & 222 & 81 & 2 \\
                                    & H1  & 1242  & 216 & 81 & 2 \\
\cmidrule{1-6}
\multirow{3}{*}{Senate\_committees}& H   & 282 & 301& 31& 2\\
                                    & H0  & 266  & 198 & 31 & 2 \\
                                    & H1  & 282  & 195 & 30 & 2 \\
\cmidrule{1-6}
\multirow{3}{*}{Senate\_bills}        & H   & 294 & 21721& 99& 2\\
                                    & H0  & 294  & 14462 & 99 & 2 \\
                                    & H1  & 293  & 14516 & 98 & 2 \\
\cmidrule{1-6}
\multirow{3}{*}{Gene\_diease}        & H   & 9262 & 3128& 1074& 25\\
                                    & H0  & 8259  & 2029 & 1074 & 25 \\
                                    & H1  & 7873  & 2043 & 1074 & 25 \\
\cmidrule{1-6}
\multirow{3}{*}{arxiv}      & H   & 111421& 172173& 2097& 10\\
                            & H0  & 111421  & 94445 & 2009 & 10 \\
                            & H1  & 111351  & 94035 & 2097 & 10 \\
\cmidrule{1-6}
\multirow{2}{*}{Gene-Protein}      & gene  & 357  & 107 & 43 & 56 \\
                                    & protein  & 988  & 1207 & 492 & 135 \\
\cmidrule{1-6}
\multirow{2}{*}{Author-Citation}& Author  & 133922  & 78805 & 115 & - \\
                                        & Citation& 124795  & 78805 & 178 & - \\
\bottomrule
\end{tabular}
\footnotetext{The table presents the hyergraph names, where `H' represents the original hypergraph, `N' indicates the number of nodes, `|E|' denotes the number of hyperedges, `D'  represents the maximum hyperedge size, and `K' represents the number of real communities.}
\end{table}



We employed two stochastic block models as baselines: 
(i) Hypergraph-MT \cite{bib10}. 
It is a probabilistic generative model for hypergraphs with mixed-membership community structure. 
It is capable of detecting  community structures in small to medium-scale hypergraphs. 
Following the processing procedure in work \cite{bib10},
we restrict the sizes of some large hyperedges in certain datasets to  enable Hypergraph-MT to run on large hypergraphs.; 
(ii) Hy-MMSBM \cite{bib19}. It is the hypergraph mixed-membership stochastic block model that mines communities from a single hypergraph.  
For each hypergraph in the multi-hypergraph, we run the baseline 10 times and retain the result with the highest likelihood; for MHSBM, we directly run it 10 times on the multi-hypergraph to obtain the highest likelihood result.

\subsubsection{The advantages of multi-hypergraph} \label{subsubsec4}
\begin{figure*}[!t]
    \centering
    \includegraphics[width=1\linewidth]{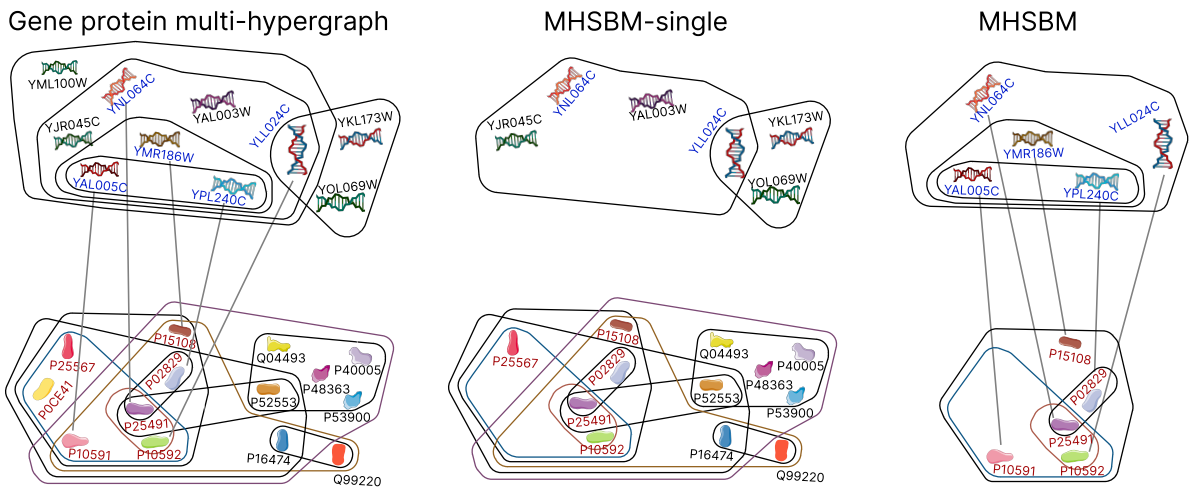}
    \caption{\textbf{The advantages of multi-hypergraph: an illustrative example.} 
    The left plot shows the partial hypergraph structure  of the protein complex HMC and its associated gene clusters within the gene-protein multi-hypergraph. 
 Gene IDs are marked in blue and protein IDs in red within the ground-truth community. All other protein and gene IDs outside this community are marked in black. 
    Gray lines represent gene-protein encoding relationships, while hyperedges are represented by irregular closed curves. 
    The central plot illustrates the protein complexes and gene clusters detected by MHSBM-single (without inter-hypergraph edge information), whereas the right displays results from inter-hypergraph informed detection. 
    This example demonstrates the advantage of multi-hypergraph modeling, where the model improves detection accuracy by leveraging complementary inter-hypergraph edge information.
}
    \label{fig1}
\end{figure*}
%
In this section, we construct a multi-domain hypergraph composed of a gene regulatory hypergraph and a protein-protein interaction hypergraph, termed as ``Gene protein multi-hypergraph". The edge between these two hypergraphs is established through gene-protein encoding relationships. 
Detailed information of this multi-domain hypergraph are provided in Table \ref{tab:datasets}. 
For the gene hypergraph, hyperedges represent gene regulatory relationships and are sourced from the Saccharomyces Genome Database (\href{https://www.yeastgenome.org/}{https://www.yeastgenome.org/}). For the protein hypergraph, hyperedges model protein-protein interactions and are constructed using interaction data from the UniProt (\href{https://www.uniprot.org/uniprotkb?query=yeast}{https://www.uniprot.org/uniprotkb?query=yeast}) and KEGG (\href{https://www.genome.jp/kegg-bin/show_organism?menu_type=pathway_maps&org=sce}{https://www.genome.jp}) databases. 
We use protein complexes from the Complex Portal (\href{https://www.ebi.ac.uk/complexportal/home}{https://www.ebi.ac.uk/complexportal/home}) biological database as ground-truth communities for the protein hypergraph, and the gene clusters formed by genes corresponding to  protein complexes as ground-truth communities for the gene hypergraph. 
For the analysis, we run MHSBM on the three hypergraphs: the gene regulatory hypergraph, protein-protein interaction hypergraph, and the Gene protein multi-hypergraph.
We constructed a case study using the HMC protein complex (UniProt IDs of the proteins: [P02829, P15108, P10591, P25567, P0CE41, P10592, P25491]) and its associated gene cluster (gene IDs: [YPL240C, YMR186W, YAL005C, YLL024C, YNL064C]). 
We employ precision and recall \cite{precisionandrecall} as evaluation metrics. 


Fig. \ref{fig1} shows the communities, corresponding to HMC protein complex, detected by MHSBM.
The performance of MHSBM on the Gene protein multi-hypergraph is superior to that of MHSBM on single hypergraphs.
On the single gene regulatory hypergraph, our model correctly detected two genes (YNL064C, YLL024C), erroneously identified four  genes (YJR045C, YOL069W, YAL003W, YKL173W), and missed three genes (YPL240C, YMR186W, YAL005C). 
In contrast, on the multi-hypergraph, the communities detected by our model exactly matched  the ground-truth gene cluster.
The complementary topological patterns captured by inter-hypergraph edges led to exact recovery of the gene cluster, achieving maximal precision and recall (both 1.0).
On the single protein hypergraph, MHSBM achieved a precision of 0.4167 (five proteins correctly detected, seven unrelated proteins erroneously detected) and a recall of 0.7142 (five  proteins detected, two proteins missed).
On the multi-hypergraph structure, our model achieved a precision of 1.0 and a recall of 0.7142 due to the omission of two proteins (P0CE41, P25567).
These results indicate that edges representing functional associations between protein interaction hypergraph and gene regulatory hypergraph \cite{robustness} improve the detection performance for both protein complexes and functional gene clusters.
This example demonstrates potential advantageous scenarios on multi-domain hypergraphs, where inter-hypergraph edge information enhances the performance of detection through cross-domain complementarity.

\begin{figure*}[!t]
    \centering
    \includegraphics[width=1\linewidth]{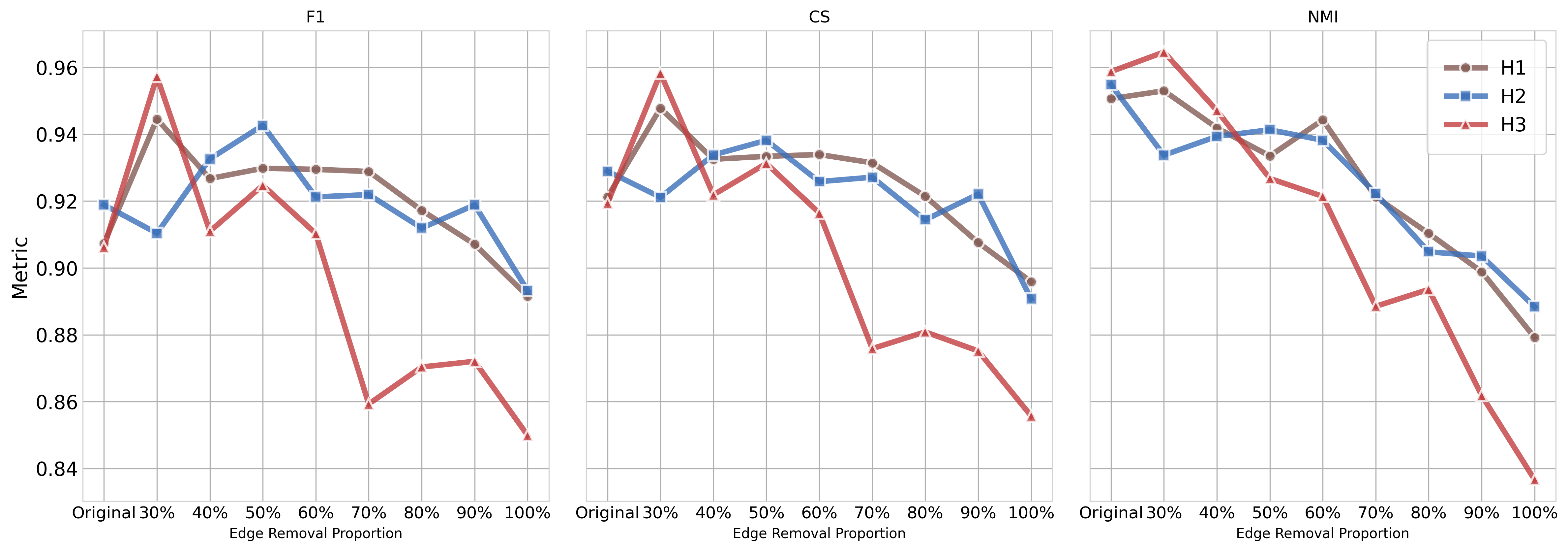}
    \caption{
    \textbf{Effect of inter-hypergraph edges on the performance of MHSBM.}
We show the metric variations of MHSBM on the multi-hypergraph across different removal proportions of inter-hypergraph edges.
The horizontal axis represents the proportion of inter-hypergraph edges removed between the hypergraph (ranging from 30\% to 100\%). 
The vertical axis shows the model's F1-score, CS, and NMI metrics for each hypergraph (H1/H2/H3).
This plot shows that the performance of MHSBM has a declining trend, with some fluctuations, as inter‑hypergraph edges decrease. 
}
  
    \label{fig2}
\end{figure*}
We now focus on investigating  the effect of the number of inter-hypergraph edges in a multi-hypergraph on the performance of MHSBM on Highschool dataset. 
Highschool dataset describes contact interactions among high school students.  
We sample 20\% hyperedges from the Highschool hypergraph three times independently to obtain three derived hypergraphs (labeled H1, H2, and H3 respectively).
We then generated inter-hypergraph edges based on the correspondence of communities in multiple hypergraphs and introduced noise perturbations to simulate real-world interference.   
Specifically, we generate 2,867 inter-hypergraph edges between H1 and H2; 2,792 edges are established between H1 and H3.

Fig. \ref{fig2} illustrates  F1-score \cite{f1, f1_2}, normalized mutual information (NMI) \cite{NMI,NMI2,NMI3}, and cosine similarity (CS) \cite{bib10,bib19} of  MHSBM under varying numbers of inter-hypergraph edges. 
It is evident that as the number of inter-hypergraph edges decreases, all indicators exhibit a declining trend. 
When we increase the removal proportion of inter-hypergraph edges to 50\%, MHSBM still maintains high performance on multi-hypergraph (e.g., a F1-score of 0.9298 on hypergraph H1).
When all inter-hypergraph edges are removed from a multi-hypergraph, MHSBM achieves an F1-score of 0.8499 on hypergraph H3, lower than the F1-score of 0.9012 on the original multi-hypergraph.
The reduction of inter-hypergraph edges hinders effective information complementarity between networks, which in turn causes the model's performance to deteriorate.
This finding demonstrates the crucial role of inter-hypergraph edges.

Through this series of experiments, we have clearly demonstrated the possible advantages of community detection in multi-hypergraph. Multi-hypergraph can effectively integrate information from each network through inter-hypergraph edges to comprehensively characterize the complex structure of the system and improve the quality of the detected communities.

\begin{figure*}[!t]
    \centering
    \includegraphics[width=0.9\linewidth]{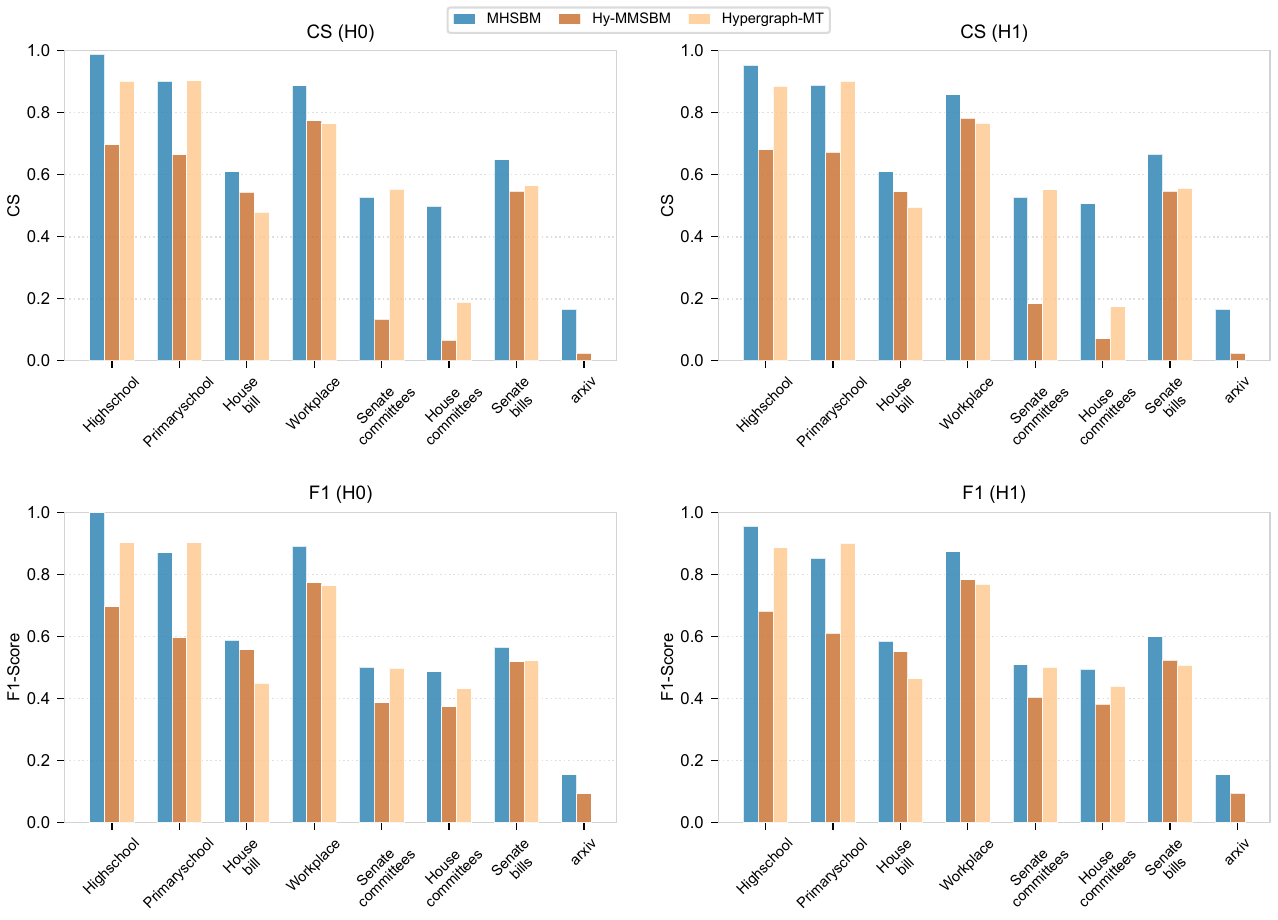}
    \caption{\textbf{Comparison of community detection algorithms on multi-hypergraphs.}
    We show the performance of MHSBM, Hy-MMSBM, and Hypergraph-MT on  multi-hypergraphs. 
For each model, seven distinct random seeds are used for initialization, resulting in seven sets of outcomes from which the mean values are computed. 
    }
    \label{fig3}
\end{figure*}
\subsubsection{Community detection on multiple  hypergraphs} \label{subsubsec5}
We selected high-order datasets from various fields (see Table \ref{tab:datasets}). 
For each dataset, we employ F1-score \cite{f1,f1_2} and CS \cite{bib10,bib19} to assess how well the inferred communities align with ground-truth distributions. 

As shown in Fig. \ref{fig3}, MHSBM outperforms  baselines on multiple datasets.
On the Workplace dataset, in  terms of CS, Hy-MMSBM achieves 0.7734 on hypergraph H0 and 0.7827 on H1, while Hypergraph-MT attains 0.7643 on H0 and 0.7651 on H1. 
Compared with  Hy-MMSBM and Hypergraph-MT,  our method demonstrates superior performance with CS of 0.8864 on H0 and 0.8584 on H1.
The superior performance of MHSBM over baseline methods can be explained from two aspects.

\begin{table}[!t]
\centering
\setlength{\tabcolsep}{5pt}
\caption{Comparison of community detection algorithms  on the Workplace dataset}    
\label{tab2:workplace}
\begin{tabular}{>{\centering}p{0.5cm} >{\centering}p{0.7cm}>{\centering}p{1.1cm}>{\centering}p{1.05cm}>{\centering}p{1cm}p{1.1cm}} 
\toprule
   &hyper graph&Hy-MM SBM&Hypergr aph-MT&MHSBM&MHSBM-single\\
\midrule
\multirow{2}{*}{NMI}   &H0 &0.6847 	&0.6854 	&0.7903 	&0.7152\\
                       &H1   &0.6831 	&0.7007 	&0.8296 	&0.7108\\

\cmidrule{1-6}
\multirow{2}{*}{\parbox{0.7cm}{\centering F1- \\ score}} &H0 &0.7726 	&0.7631&0.8913 	&0.8010\\
                    &H1 &0.7834&0.7686&0.8731 	&0.8043\\
\cmidrule{1-6}
\multirow{2}{*}{CS} &H0 &0.7734 	&0.7643&0.8864 	&0.7921\\
                    &H1 &0.7827 	&0.7651&0.8584 	&0.7976\\
\bottomrule
\end{tabular}
\footnotetext{
We present  the NMI, F1-score, and CS of algorithms on H0 and H1, respectively. 
MHSBM-single denotes the model MHSBM that utilizes only single-hypergraph structural information.
}
\end{table}
First, the MHSBM model leverages  information from multiple hypergraphs. 
In contrast, Hypergraph-MT and Hy-MMSBM mainly rely on information extracted from a single hypergraph, failing to utilize the information from other hypergraphs.
To validate that the MHSBM  improves community detection performance by leveraging information from multiple hypergraphs,  
we compared the performance of the MHSBM on single hypergraphs and on multiple hypergraphs of  the Workplace dataset,  reporting  the results in Table \ref{tab2:workplace}.
We refer to the MHSBM applied to single hypergraphs as `MHSBM-single'. 
The experimental results demonstrate that MHSBM exhibits significantly superior performance compare with the MHSBM-single.
On H0, MHSBM achieves an F1-score of 0.8913, demonstrating an improvement over MHSBM-single (0.8010).
Similarly, the CS and NMI of MHSBM on multi‑hypergraphs exceed those of MHSBM‑single.

Second, `hyperedge internal degree' introduced by our model could accurately  model the hyperedge, as evidenced by  MHSBM-single outperforming the baselines.
The idea of `hyperedge internal degree' is based on a phenomenon on the Workplace dataset.
The conditional probability of hyperedges of size 2 being contained within hyperedges of size greater than 2 equals 1. It  indicates a strong correlation between the formation of small and large hyperedges. 
We also compute information entropy \cite{infomation_entroy1} of hyperedges  to explore the distribution characteristics of hyperedges.
The  probability of each node in hyperedge $e$ is   the  number of hyperedges that contain this node and are the subset of $e$, divided by the sum of such counts for all nodes in $e$.
Approximately 10\% of hyperedges have information entropy values below 0.6.
This intuitively reflects non-uniform distribution  of small hyperedges within large hyperedges. 
It may stem from the preferential attachment mechanism \cite{preferential-attachment-mechanism}, wherein high-importance nodes exhibit heightened propensity to attract new connections. 
This observation empirically substantiates the design logic of hyperedge internal degree.
This also allows MHSBM to place different emphasis on the nodes to the formation of hyperedges, making our model more aligned with real-world scenarios. 

\begin{figure*}
    \centering
    \includegraphics[width=1\linewidth]{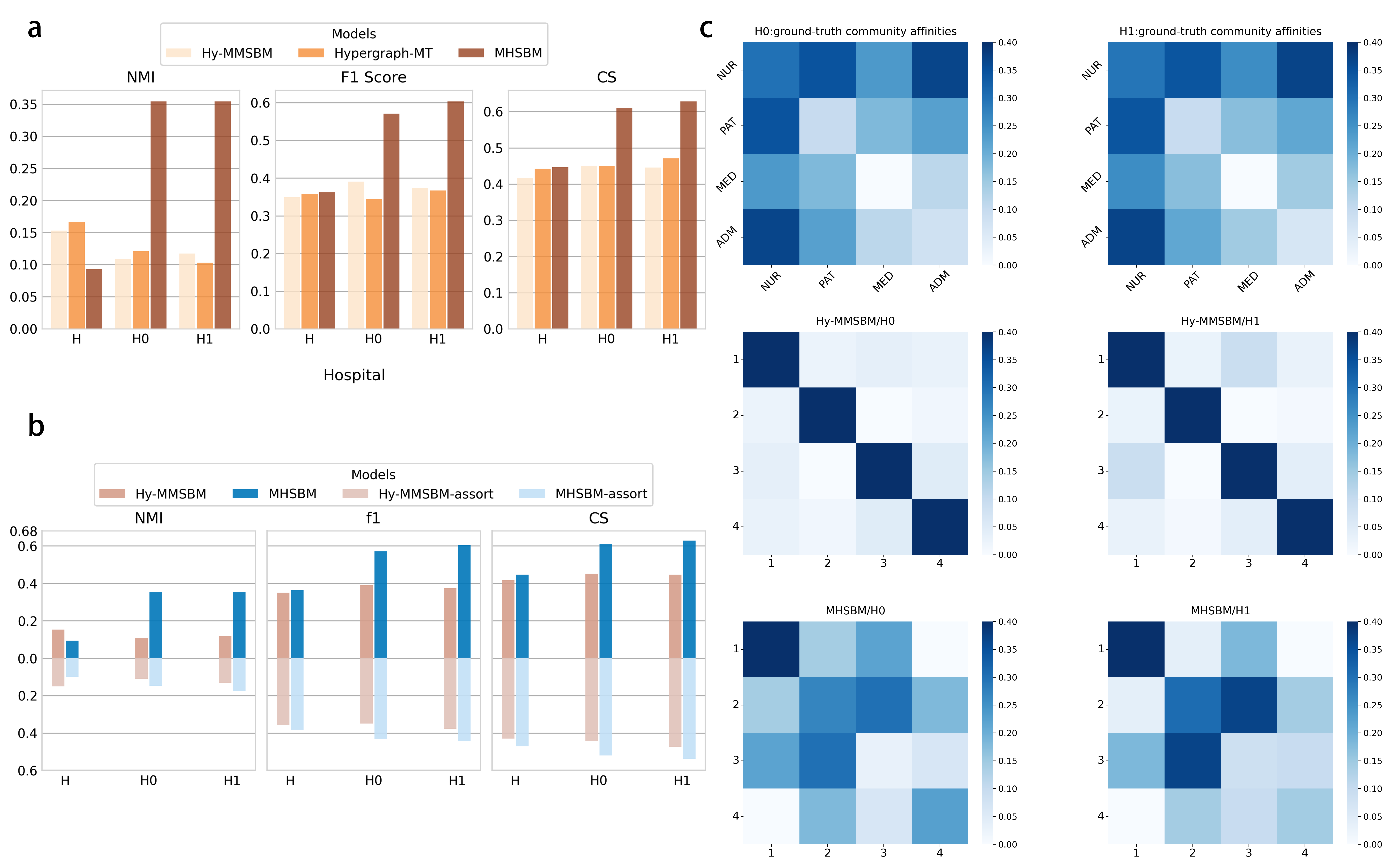}
    \caption{
    \textbf{Community configuration on the hospital dataset.} 
\textbf{a} Metrics of MHSBM, Hy-MMSBM, and Hypergraph-MT. For MHSBM and Hy-MMSBM models, we set the assortative to false.
    \textbf{b} Results of MHSBM and Hy-MMSBM under different initializations of  $w$. 
    Results above the horizontal axis are the detection results with the assortative parameter set to False, and  those below the axis are the results with assortative set to True.
    \textbf{c}
    Community affinities.
    The first row of plots shows the community affinities of the true communities in H0 and H1, the second row displays the community affinities captured by  Hy-MMSBM, and the third row shows the community affinities captured by  MHSBM.
    }
    \label{fig4}
\end{figure*}

Then, we further analyzed the ability of the MHSBM model to detect community configurations on the Hospital dataset.  
In this dataset, nodes represent patients and healthcare workers.   Hyperedges capture three types of interactions: interactions between patients, interactions between patients and healthcare workers, and interactions between healthcare workers.
The nodes are divided into four communities: nurses and nurse assistants, patients, doctors, and administrative staff. The dataset exhibits a clear disassortative community structure.
Fig. \ref{fig4} shows the performance of MHSBM and baseline methods on the Hospital dataset.  
Particularly, on the multi-hypergraph, MHSBM achieves the best performance across all metrics (Fig. \ref{fig4}a), with CS values of 0.6104 and 0.6280 on H0 and H1 respectively. 

To evaluate the ability of MHSBM to capture disassortative community structures, we compared the detection results under different initializations of the affinity matrix $w$.  
Specifically, if assortative is set to True, all off-diagonal elements of \( w \) are initialized to zero; if assortative is set to False, non-zero values are assigned to the off-diagonal elements of \( w \).
Fig. \ref{fig4}b shows the performance of Hy-MMSBM and MHSBM on the Hospital dataset under assortative = True and assortative = False. 
MHSBM with the assortative parameter set to False performs significantly better than MHSBM with it set to True. 
On H0, MHSBM's CS is 0.5209 when assortative=True and 0.6104 when assortative=False.
Other metrics also display similarly significant improvements (Fig. \ref{fig4}b).
This significant improvement arises from MHSBM's sensitivity to different community configurations (assortative and disassortative structures ) \cite{bib19}. 
In contrast, Hy-MMSBM exhibits no significant variation under the two initialization.
For instance, on H0, Hy-MMSBM's CS is 0.4501 when assortative=True and 0.4426 when assortative=False (as shown in Fig. \ref{fig4}b).
To further analyze the above result, Fig. \ref{fig4}c presents the affinities between true communities in the multi-hypergraph (first rows),  the community affinities captured by Hy-MMSBM   (second row) and those identified by MHSBM (third row). Here, the assortative parameters of both Hy‑MMSBM and MHSBM are set to False. 
As shown in the Fig. \ref{fig4}c, the ground-truth communities exhibit a clear disassortative structure, with strong interactions between communities and relatively sparse interactions within communities.
The community affinities captured by our model reveal  strong connections between two communities, 
while the Hy-MMSBM model fails to capture the dense interaction between communities.
Compared with Hy-MMSBM, the community affinities captured by the MHSBM model are more similar to the true community affinities, further validating the model's ability to capture community configurations in multi-hypergraphs. 
These experimental results validate the reliability of the MHSBM model in community detection tasks. 
\subsubsection{Hyperedge prediction within the hypergraph} \label{subsubsec6}
In this section, we focus on the model's ability to predict missing hyperedges  within hypergraphs and adopt cross-validation to select the optimal parameters. 
We analyzed a high-order gene-disease dataset, in which nodes represent genes and hyperedges consist of genes associated with diseases. The gene-disease  dataset are divided into training and test sets. The training set is utilized for model parameter estimation, and the test set is employed to evaluate the model's performance in predicting nonobserved hyperedges.
We use the Area Under the Curve (AUC) \cite{bib10,bib19} as the evaluation metric. 
The AUC value ranges from 0 to 1; a higher AUC indicates more accurate predictions by the model.

\begin{figure}[!t]
    \centering
    \includegraphics[width=1\linewidth]{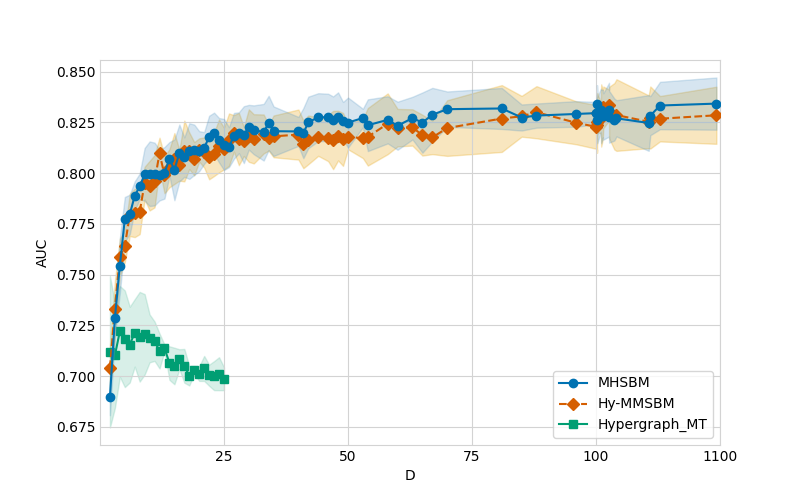}
    \caption{\textbf{Hyperedge prediction on the Gene-Disease dataset}. 
    We evaluate the prediction performance by the mean and standard deviation of the AUC under different maximum hyperedge sizes D in the test set.
    The line plots depict the AUC variation trends of  methods, with shaded regions indicating the standard deviation of AUC values.
    The figure illustrates the stability of the MHSBM for hyperedge prediction at varying hyperedge sizes.}
    \label{fig5}
\end{figure}

For the baseline method, we conduct hypergraph prediction trials for 10 times on each hypergraph in the multi-hypergraph. 
We calculate the model's AUC for each trial. For each hypergraph, we compute the mean and standard deviation of the AUC values across 10 trials.
The model's overall performance is evaluated by averaging the mean AUC values (and their standard deviations) across all hypergraphs in gene-disease dataset.
We evaluate the predictive performance of methods under different values of the maximum hyperedge size D.
Our method achieves AUC scores between 0.68 and 0.834 across different values of D.
As shown in Fig. \ref{fig5}, when D = 1071, our method attained an AUC score of 0.8342, outperforming Hy-MMSBM with an AUC of 0.8285. 
Overall, our method demonstrates solid performance in predicting hyperedges of different sizes. 

We further extended our analysis to various datasets from different domains. 
For each dataset, we report the mean and standard deviations of AUC score for both the MHSBM and Hy-MMSBM in Table \ref{tab3}. 
\begin{table}
\centering
\caption{AUC comparison of algorithms for predictive performance}
\label{tab3}
\begin{tabular}
{p{0.25\columnwidth} p{0.30\columnwidth} p{0.30\columnwidth}}
\toprule
Dataset&MHSBM&Hy-MMSBM\\
\midrule
Highschool & $0.9225\pm 0.008$& $0.9213 \pm 0.008$\\
\midrule
Primary & $0.8846 \pm 0.010$ & $0.8573 \pm 0.008$\\
\midrule
Hospital  & $0.7984\pm 0.009$&$0.7576 \pm  0.012$\\
\midrule
Housebill & $0.9526\pm  0.007$& $0.9243 \pm  0.005$\\
\midrule
Workplace& $0.7680\pm  0.010$ & $0.7368 \pm  0.018$\\
\midrule
Senatebills& $0.931 \pm 0.007$ & $0.926 \pm 0.006$ \\ 
\bottomrule
\end{tabular}
\footnotetext{We present the mean  and standard deviation of AUC for hyperedges achieved by the MHSBM and Hy-MMSBM models across various datasets.}    
\end{table}
The AUC scores of MHSBM range from 0.7680  to 0.9526, while those of Hy-MMSBM range from 0.7368  to 0.9243. 
This indicates that MHSBM exhibits strong fitting capability and can accurately predict the presence of hyperedges. 
\enlargethispage{2\baselineskip} 
The enhanced performance is attributed to our model's integration of information from multiple hypergraphs to optimize prediction accuracy.

In summary, these results indicate that our model exhibits excellent performance in handling high-order datasets from different domains.

\subsubsection{Inter-hyergraph edge prediction} \label{subsubsec7}
Our model is  capable of predicting interactions between hypergraphs, i.e., inter-hypergraph edges.
Here, we analyze a dataset called Author-Citation, composed of the author hypergraph and the paper hypergraph. Hyperedges and inter-hypergraph edges in the Author-Citation dataset are constructed using data from the Aminer (\href{https://www.aminer.cn/citation}{https://www.aminer.cn/citation}) database.
Author-Citation dataset contains  three types of interactions: (1) co-authorship relationships among authors in  the author hypergraph, (2) citation relationships  between  papers in the paper hypergraph, and (3) author-citation relationships associating researchers with their papers  between the  author hypergraph and the paper hypergraph. To evaluate the effect of inter-hypergraph edges on MHSBM's prediction performance, we vary the ratio $r$ of inter-hypergraph edges removed. The ratio $r$  increased from 0.1 to 0.9 with a step size of 0.1. The remaining inter-hypergraph edges were partitioned into training and test sets at a 4:1 ratio. Then, we calculated the AUC value  \cite{bib10,bib19} at each value of $r$. 

\begin{figure} 
    \centering
    \includegraphics[width=1\linewidth]{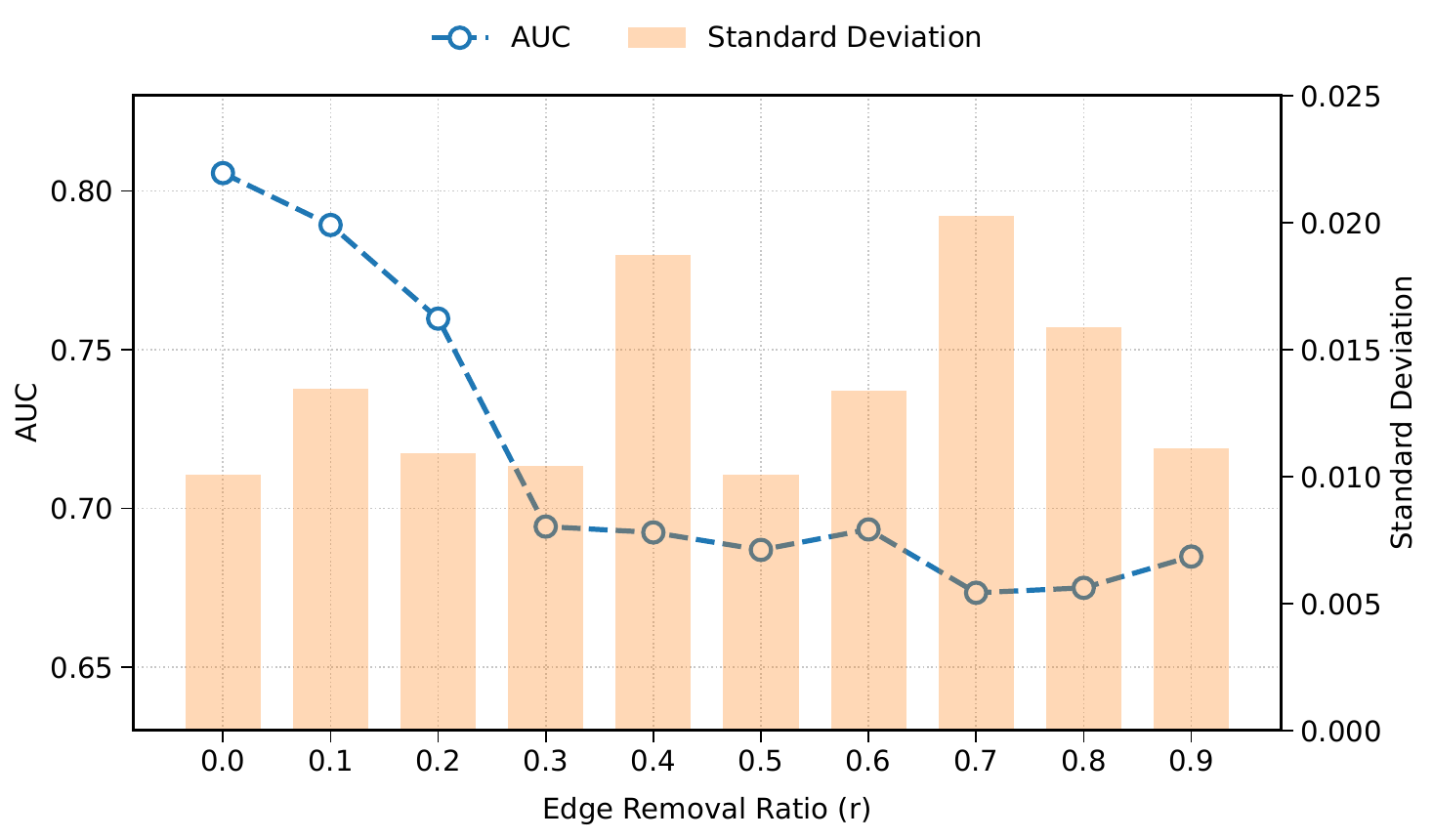}
    \caption{\textbf{\textbf{Inter-hypergraph edge prediction by MHSBM on the Author-Citation dataset} }The blue dashed line represents the AUC variation curve, and the orange bars indicate the standard deviation. Here, removal ratio denotes the proportion of inter-hypergraph edges removed from the training set. When \(r = 0\), the AUC is \(0.8056 \pm0.01008\); at \(r = 0.7\), the AUC reaches its minimum value.
    }
    \label{fig6}
\end{figure}

%
As shown in Fig. \ref{fig6}, when the  ratio $ r=0$ (i.e., no inter-hypergraph edges were removed), the model achieved the optimal AUC of $0.8056 \pm 0.01008$. 
As $r$ gradually increased, the model performance exhibited a nonlinear degradation trend. 
Specifically, for $ r \le 0.2$, with the increase of ratio $r$, the AUC declined mildly to $0.7598 \pm 0.01093 $. 
At $r=0.3$, a significant performance inflection point emerged: compared with the AUC of $0.7598 \pm 0.0081$ at r=0.2, the AUC at r=0.3 exhibited a sharp decline of 6.55\% to $0.6943 \pm 0.01042$.
This result indicates that the removal of inter-hypergraph edges hinders information interaction between hypergraphs, thereby resulting in a progressive degradation of the model’s predictive performance.
Notably, when the removal ratio $r \ge 0.4$, the model maintained relatively stable predictive performance (AUC range: 0.6734–0.6934), indicating reduced sensitivity to edge removal at this stage. Even when $r=0.9$, where only a small fraction of inter-hypergraph edges remained, our model retained basic predictive capability through the surviving connections.
The experimental results demonstrate that the model's predictive performance is influenced by the number of inter-hypergraph edges, with fewer edges removed leading to better prediction performance.

Our model was further validated across diverse domain datasets, with experimental results shown in Table \ref{tab4}. 
The achieved AUC scores ranging from 0.7734 to 0.9643 confirm creliable predictive performance across diverse datasets. 
This capability enables the model to capture latent multiple high-order interactions, thereby advancing analytical precision in modeling complex real-world networked systems.

\begin{table}
\centering
\caption{Inter-hypergraph edge prediction performance of d MHSBM on high-order datasets }
\label{tab4}
\begin{tabular}
{p{0.23\columnwidth} p{0.1\columnwidth} p{0.29\columnwidth} p{0.1\columnwidth}}
\toprule
dataset & AUC  & dataset&AUC\\ 
\midrule
Highschool & $0.9643 $& Diease&$0.7734$\\ 
\midrule
Housebills&  $0.8655 $& Hospital&$0.8097 $\\ 
\midrule
Primary& $0.9718 $& Workplace&$0.8262$\\
\midrule
 Senatebills& $0.9170 $& Author-Citation&$0.8056$\\
 \bottomrule
\end{tabular}
\footnotetext{It presents the  values and their  standard deviations for inter-hypergraph edge prediction on various multi-hypergraphs.
}
\end{table}
\section{Discussion} \label{sec3}
We investigated the problem of  community detection in multi-hypergraphs and proposed a probabilistic generative model, named MHSBM.
To the best of our knowledge, this is the first work to identify community  from  multiple high-order network systems. 
In addition to detecting communities in the multi-hypergraph, our model can also predict missing hyperedges within a hypergraph and missing edges between hypergraphs. 
MHSBM integrates the structural information across  hypergraphs, enabling the transfer of information between distinct hypergraphs, and leverages this information to extract the organizational structure of nodes.
In the case analysis of the Hospital dataset, we demonstrate the  model's ability to detect community configurations. 
Furthermore, our model performs hyperedges prediction both within hypergraphs and across multiple hypergraphs,
demonstrating improved performance in predicting missing hyperedges by leveraging the  complementary information between hypergraphs.  

Our model is characterized by two key features.
Firstly, the introduction of the  community affinity matrix across hypergraphs enables our model to facilitate information complementarity between different high-order systems.
Ignoring this information leads to a performance decline in the experiments, as shown by the analyses of community detection and hypergraph prediction tasks.
It is worth noting that our model is applicable to multi-view hypergraphs that describe the relations of the same nodes from different perspectives; as well as to  multi-domain hypergraphs, where nodes and edges are distinct across different domains.
Secondly, we also performed community detection on a single hypergraph, demonstrating our model's strong community recovery capability even in single-hypergraph structures. This is attributed to the introduction of the hyperedge internal degree matrix, which treats the contributions of different nodes within a hyperedge differently.

There are several potential extensions for our model. 
Our model focuses on the modeling of pairwise interactions between hypergraphs.
It might be generalized to support  high-order interactions among  hypergraphs, a critical feature of  complex high-order multi-systems. 
On a related note, our model integrates complementary information across multiple hypergraphs by utilizing all inter-hypergraph edges. 
The complementary information  does not always produce positive effects. For example, a hypergraph may provide effective complementary information to other hypergraphs, or it may create interference for them. 
Our model might be further extended to adaptively leverage beneficial information  from specific hypergraphs while suppressing interference.
Finally, our model might be extended to temporal hypergraph analysis to investigate dynamic systems.
Overall, our work provides a reliable, flexible, and scalable tool for the modeling of multi-hypergraph, enhancing our ability to handle and study the organization of real-world high-order multi-systems.

\section{Methods} \label{sec4}
\subsection{Inference of MHSBM} \label{subsec4}
Based on  Eqs. (\ref{eq2}) and (\ref{eq4}), MHSBM models the hyperedges within hypergraphs and the edges between hypergraphs as follows:
\begin{footnotesize}
\begin{equation}
\begin{aligned}
        P(A,S|\Phi) &=\prod_{l \in [0,L]} \prod_{e \in \Omega^l} Pois(A_{e}^{l};~\frac{\lambda_{e}^{l}}{\mu_{e}^{l}})~\times~\\
        &\prod_{l^{\prime} \ne l:l,l^{\prime}\in [0,L]}~ \prod_{\substack{i \in  V^l,j \in  V^{l^{\prime}}}}Pois(S_{ij}^{l l^{\prime}};\lambda_{ij}^{l l^{\prime}}) , 
\end{aligned}
    \end{equation}
    \end{footnotesize}where $\Omega^l$ represents the space of all possible hyperedges. The large size of the hyperedge space makes parameter updating extremely complex. Previous work \cite{bib10,bib19} points out that the space $\Omega^l$ can be reduced. 
Therefore, in this paper, we use \( E^{l+} \cup E^{l-} \) to replace \( \Omega^l \), where \( E^{l+} \) is the set of observed hyperedges, and \( E^{l-} \) is the set of possible hyperedges sampled from \( \Omega^l - E^{l+} \), with the same size as \( E^{l+} \).
For each pair of corresponding hyperedges $e^{l+} \in E^{l+}$ and $e^{l-} \in E^{l-}$, their sizes (number of nodes) are equal, i.e., $|e^{l+}| = |e^{l-}|$.

Then, we use the Maximum Likelihood Estimation (MLE) method to infer the parameters, with the likelihood function given by: 
\begin{footnotesize}
  \begin{equation}
    \begin{aligned}
    &\log p(A, S|\Theta) =\\
    & \underbrace{\sum_{l \in [0,L]} [\sum_{\substack{e\in \\ E^{l+}\cup E^{l-} } }-\frac{1}{\mu_{e}^l}\sum_{i<j \in e} t+ \sum_{\substack{e\in \\ E^{l+}\cup E^{l-} }} A^l_{e} \mathrm{log} \sum_{i<j\in e} t]}_{L_A}+ \\ 
    & \underbrace{ \sum_{l \ne l^{\prime}:l,l^{\prime} \in [0,L]}[-\sum_{\substack{i \in  V^l,\\ j \in  V_{l^{\prime}}} }u_i^{l} w^{ll^{\prime}} {u_j^{l^{\prime}}}^T+  \sum_{\substack{i\in  V^l,\\j \in  V_{l^{\prime}}} }S_{ij}^{l l^{\prime}} \mathrm{log} \sum_{k, c} u_{ik}^{l}w_{kc}^{l l^{\prime}}u^{l^{\prime}}_{jc}] }_{L_S},
    \label{eq6}
    \end{aligned}
    \end{equation}
    \end{footnotesize}where $t$ represents $\theta_{ie}\theta_{je} {u_{i}^l} w^l {u^l_{j}}^T$.
Here, we ignore the constant terms in \( L_A \) and \( L_S \) that do not depend on the parameters. 
The term  \( \sum_{e \in E^{l+} \cup E^{l-}} -\frac{1}{\mu_{e}^l} \sum_{i<j \in e} \theta_{ie} \theta_{je} {u_{i}^l} w^l {u^l_{j}}^T \) in \( L_A \) requires the calculation of the interaction strength between every pair of nodes in all hyperedges of \( E^{l+} \cup E^{l-} \).  The computation of this term faces high computational cost when the hyperedge size or the number of hyperedges is large.  
 To compute efficiently, we approximate this term by multiplying the average interaction strength between all node pairs in the hypergraph  by the number of node pairs in all hyperedges. 
 Specifically, the average interaction strength of \( E^{l+} \) is \( \frac{\sum_{i < j \in  V^l} u_{i}^{l} w^{l} {u_{j}^{l}}^T}{|Q|} \), where \( Q \) is the number of node pairs in  hyperedges in \( E^{l+} \). 
For node pairs without an edge, the value of \( u_{i}^{l}w^{l} {u_{j}^{l}}^T \) is close to zero, so its effect is negligible. 
The the average interaction strength in \( E^{l-} \) is \( \frac{\sum_{i < j \in  V^l} u_{i}^{l} w^{l} {u_{j}^{l}}^T }{(| V^l|(| V^l|-1))/2}\), where \( \frac{| V^l|(| V^l|-1)}{2} \) is the number of binary interactions.
The number of node pairs in 
\( E^{l+} \)  and  \( E^{l-} \)  is  \( \sum_{e \in E^{l+}} \frac{|e|(|e|-1)}{2} \) and \( \sum_{e \in E^{l-}} \frac{|e|(|e|-1)}{2} \),  respectively.
Furthermore, we simplify the hyperedge internal degree \( \theta \) to 1. 
The term  $\sum_{e \in E^{l+} \cup E^{l-}} -\frac{1}{\mu_{e}^l} \sum_{i<j \in e} \theta_{ie}\theta_{je} {u_{i}^l} w^l {u^l_{j}}^T $ is approximated as:
\enlargethispage{3\baselineskip} 
\begin{footnotesize}
\begin{equation}
    -M~ \left(\frac{1}{|Q|} + \frac{1}{\frac{| V^l|(| V^l|-1)}{2}}\right) \cdot  \sum_{i < j \in  V^l} {u_{i}^{l}} {w}^{l} {u_{j}^{l}}^T. \nonumber
\end{equation}
\end{footnotesize}After the simplification, the final approximate result for the term 
$\sum_{e \in \Omega^l } -\frac{1}{\mu_{e}^l} \sum_{i<j \in e} \theta_{ie}\theta_{je} u_{i}^l w^l (u_{j}^l)^T$ is obtained as: 
\begin{footnotesize}
\begin{equation}
\begin{aligned}
   \sum_{e \in \Omega^l} -\frac{1}{\mu_{e}^{l}} \sum_{i < j \in e} \theta_{ie} \theta_{je} u_{i}^{l} {w}^{l} {u_{j}^{l}}^T
    = C_l \cdot  \sum_{i < j \in  V^l} {u_{i}^{l}} {w}^{l} {u_{j}^{l}}^T, 
\end{aligned}
\end{equation}
\end{footnotesize}where \( C_l = -M\left( \frac{1}{|Q|} + \frac{1}{\frac{|V^l|(|V^l|-1)}{2}} \right) \).  
The Eq. (\ref{eq6})  is thereby simplified as:
  \begin{footnotesize}
  \begin{equation}
    \begin{aligned}
    &\log p(A, S|\Theta) =\\
   &\sum_{l \in \mathcal{L}} [-C_l \sum_{i<j \in  V^l}  u_{i}^l w^l {u^l_{j}}^T+\\ &\sum_{\substack{e\in \\ E^{l+}\cup E^{l-} }} A^l_{e} \mathrm{log} \sum_{i<j\in e} t] \\ 
    & + \sum_{l \ne l^{\prime}:l,l^{\prime} \in \mathcal{L}}[-\sum_{\substack{i \in  V^l,\\ j \in  V_{l^{\prime}}} }u_i^{l} w^{ll^{\prime}} {u_j^{l^{\prime}}}^T+  \\
    &\sum_{\substack{i\in  V^l,\\j \in  V_{l^{\prime}}} }S_{ij}^{l l^{\prime}} \mathrm{log} \sum_{k, c=1} u_{ik}^{l}w_{kc}^{l l^{\prime}}u^{l^{\prime}}_{jc}]. \\
    \label{eq7}
    \end{aligned}
    \end{equation}
\end{footnotesize}To maximize the likelihood function in Eq. (\ref{eq7}), we use Jensen's inequality \( \log \mathbb{E}[x] \geq \mathbb{E}[\log x] \) to obtain a lower bound for the second term in \( L_A \) and \( L_S \), respectively: 
    \begin{footnotesize}
    \begin{equation}
    \begin{aligned}
        &\sum_{e \in E^l} A^l_{e} \mathrm{log} \sum_{i<j\in e} \theta_{ie}^l  u_{i}^l w^l \theta_{je}^l {u^l_{j}}^{T}\\
       & \ge 
        \sum_{e \in E^l} A^l_{e} \sum_{i < j \in e} \sum_{k,q} p_{ijkq}^{l(e)} \log\left(\frac{\theta_{ie}^l \theta_{je}^l u_{ik}^{l} u_{jq}^{l} w_{kq}^{l} }{p_{ijkq}^{l(e)}} \right),  \label{eq8}
    \end{aligned}
    \end{equation}
    \end{footnotesize}
    \begin{footnotesize}
    \begin{equation}
    \begin{aligned}
        &\sum_{i \in V^l, j \in V^{l^{\prime} }} S_{ij}^{ll^{\prime}} log\sum_{k ,c }   u_{ik}^{l} u_{jc}^{l^{\prime}} w_{kc}^{l l^{\prime}}\\  
        &\ge \sum_{i \in V^l, j \in V^{l^{\prime} } } S_{ij}^{ll^{\prime}} \sum_{k,c } \rho_{ijkc}^{ll^{\prime}} \mathrm{log}\left( \frac{u_{ik}^{l} u_{jc}^{l^{\prime}} w_{kc}^{l l^{\prime}}}{\rho_{ijkc}^{ll^{\prime}}} \right),  \label{eq9}
     \end{aligned}
    \end{equation}
    \end{footnotesize}
where   variational distributions for Eqs. (\ref{eq8}) and (\ref{eq9}) are specified by strictly positive probabilities \( p^{l(e)}_{ijkq} \) and \( \rho_{ijkc}^{ll'} \) respectively, which satisfy the conditions \( \sum_{i<j \in e} \sum_{k,q}p_{ijkq}^{l(e)} = 1 \) and \( \sum_{k,c} \rho_{ijkc}^{ll'} = 1 \) respectively.
  
The equalities in Eqs. (\ref{eq8}) and (\ref{eq9}) hold when the following conditions are met:
    \begin{footnotesize}

    \begin{equation}
    \begin{aligned}
        p_{ijkq}^l & = \frac{\theta_{ie}^l \theta_{je}^l u_{ik}^l u_{jq}^l w_{kq}^l}{\sum_{i < j \in e} \sum_{k,q=1}^K \theta_{ie}^l \theta_{je}^l u_{ik}^l u_{jq}^l w_{kq}^l } \\
        &= \frac{\theta_{ie}^l \theta_{je}^l u_{ik}^l u_{jq}^l w_{kq}^l}{\lambda_e^l},  \label{eq10}
    \end{aligned}
    \end{equation}
\end{footnotesize}

\begin{footnotesize}
    \begin{equation}
        \rho_{ijkc}^{l l^{\prime}} = \frac{u_{ik}^l u_{jc}^{l^{\prime}} w_{kc}^{l l^{\prime}} }{\sum_{k,c=1} u_{ik}^l u_{jc}^{l^{\prime}} w_{kc}^{l l^{\prime}} }
        =\frac{u_{ik}^l u_{jc}^{l^{\prime}} w_{kc}^{l l^{\prime}} }{\lambda_{ij}^{l l^{\prime}} }.  \label{eq11}
    \end{equation}
\end{footnotesize}Therefore, maximizing $\log p(A, S|\Phi)$ is equivalent to maximizing the following expression:
      \begin{footnotesize}
    \begin{equation}
    \begin{aligned}
  & \mathcal{L}(u,w, w^{\prime},p,\rho)  = \sum_{l \in [0,L]} [-C_l \sum_{i<j \in V^l}  u_{i}^l  w^l {u^l_{j}}^T+ \\
    &\sum_{e \in E^l} A^l_{e} \sum_{i <j \in e} \sum_{k,q=1} p^{l(e)}_{ijkq} \mathrm{log} \left(\frac{\theta_{ie}^l \theta_{je}^l u_{ik}^l u_{jq}^l w_{kq}^l}{p^{l(e)}_{ijkq}} \right)]\\
    &+ \sum_{l \ne l^{\prime}:l,l^{\prime} \in [0,L]}[-\sum_{\substack{i \in V^l, \\j \in V^{l^{\prime}}} }u_i^{l} w^{ll^{\prime}} {u_j^{l^{\prime}}}^T\\
    &+ \sum_{\substack{i \in V^l,\\ j \in V^{l^{\prime}}} } S_{ij}^{ll^{\prime}} \sum_{k,c} \rho_{ijkc}^{ll^{\prime}} \mathrm{log}\left( \frac{u_{ik}^{l} u_{jc}^{l^{\prime}} w_{kc}^{l l^{\prime}}}{\rho_{ijkc}^{ll^{\prime}}} \right)].  \label{eq12}
    \end{aligned}
    \end{equation}
\end{footnotesize}The update rule for the membership vector is derived by setting the partial derivative $\partial \mathcal{L}/\partial u$ to 0: 
 \begin{footnotesize}
\begin{equation}
u_{ik}^{l} = 
    \frac{
        \sum\limits_{\substack{e \in E^l \\ i \in e}} A_{e}^{l} p_{ik}^{l(e)} 
        + 
        \sum\limits_{\substack{l' \in [0,L], \\ l' \neq l}} 
        \sum\limits_{j \in V^{l'}} 
        S_{ij}^{l l'} 
        \sum\limits_{c=1}^{C} \rho_{ijkc}^{l l'}
    }{
        C_l 
        \sum\limits_{q} w_{kq}^l 
        \sum\limits_{\substack{j \in V^l \\ j \neq i}} u_{jq}^{l} 
        + 
        \sum\limits_{\substack{l' \in [0,L], \\ l' \neq l}} 
        \sum\limits_{c} w_{kc}^{l l'} 
        \sum\limits_{j \in V^{l'}} u_{jc}^{l'}
    }. 
\label{eq13}
\end{equation}
    \end{footnotesize}For the first term in the numerator of Eq. (\ref{eq13}), we exploit the sparsity of the association matrix to reduce the computational complexity by focusing only on the non-zero values, thereby improving computational efficiency. Similarly, for the second term, we further reduce the computational complexity by utilizing the sparsity of \( S^{ll'} \).

Likewise, we can derive the update rules for the affinity matrix within the hypergraph and the affinity matrix between hypergraphs:
   \begin{equation}
        w_{kq}^{l}=\frac{\sum_{e\in E^l} A_e^{l} p_{kq}^{l(e)} }{C_l \sum_{i<j\in V^l}u_{ik}^{l} u_{jq}^{l} },  \label{eq14}
    \end{equation}
     \begin{equation}
        w^{l l^{\prime}}_{kc} =\frac{\sum_{i\in E^l,j \in E^{l^{\prime}} } S_{ij}^{l l^{\prime}} \rho_{ijkc}^{l l^{\prime}} } {\sum_{i\in E^l,j \in E^{l^{\prime}} }u_{ik}^{l} u_{jc}^{l^{\prime}} }.   \label{eq15}
    \end{equation}
Similar to the update of \( u_{ik}^l \), we fully exploit the properties of sparse matrices to improve computational efficiency.

We use the Expectation-Maximization (EM) algorithm to estimate the parameters until convergence of Eq. (\ref{eq15}). However, the EM algorithm only guarantees a local maximum, not a global maximum \cite{EMshort}. Therefore, we initialize \( \Phi \) differently and run the algorithm 10
times, selecting the maximum likelihood value and obtaining the inferred parameters.

\section*{Data availability} 
The datasets used in this paper are publicly available from their original sources. Gene-Protein and Author-Citation datasets can be downloaded from  \href{https://github.com/magical-seven/MHSBM/blob/2a0e3fb84b742fa87b7496cbbbf1c0a3f6b8c65b/dataset.zip}{https://github.com/magical-seven/MHSBM}. The remaining datasets are accessible via  \href{https://edmond.mpg.de/dataset.xhtml?persistentId=doi:10.17617/3.HRW0OE&version=1.0}{https://edmond.mpg.de/dataset.x html?persistentId=doi:10.17617/3.HRW0OE\&ve rsion=1.0}.

\section*{Code availability} 
An open-source algorithmic implementation of the model is publicly
available and can be found at \href{https://github.com/magical-seven/MHSBM.git}{https://github.com/magical-seven/MHSBM.git}.

\bibliography{sn-bibliography}

\section*{Acknowledgements} 
This work was supported by the National Natural Science Foundation of China [No.62106004, No.62272001, and No.62206004]. 

\section*{Author contributions} 
L.N. and Z.Q.D. planned and performed this research,  and
 wrote this paper. Y.W.Z., W.J.L, L.M., and L.Z. planned the research and edited this paper. All authors discussed the results and reviewed the manuscript.

\section*{Competing interests} 
The authors declare no competing interests.
\end{document}